\documentclass[aps,prl,twocolumn,showpacs,superscriptaddress,longbibliography,floatfix]{revtex4-2}
\usepackage{amsmath,amssymb}
\usepackage{graphicx}
\usepackage[utf8]{inputenc}
\usepackage{hyperref}
\usepackage{orcidlink}
\usepackage{bm}
\usepackage{xcolor}

\newcommand{\arc}{$\alpha$-RuCl$_{3}$}

\usepackage{pdfpages}
\makeatletter
\AtBeginDocument{\let\LS@rot\@undefined}
\makeatother

\begin{document}
	
	\title{Tilted and Twisted Magnetic Moments in the Kitaev Magnet \texorpdfstring{$\alpha$-RuCl$_3$}{alpha-RuCl3}}
	
	\author{Xiao Wang}
	\affiliation{Institute of High Energy Physics, Chinese Academy of Sciences (CAS), Beĳing 100049, China}
	\affiliation{Spallation Neutron Source Science Center, Dongguan 523803, China}
	\affiliation{J\"ulich Centre for Neutron Science JCNS at MLZ, Forschungszentrum J\"ulich, Lichtenbergstr. 1, D-85747 Garching, Germany}
	
	\author{Fengfeng Zhu}
	\affiliation{State Key Laboratory of Materials for Integrated Circuits, 2020 X-Lab, Shanghai Institute of Microsystem and Information Technology, Chinese Academy of Sciences, Shanghai 200050, China}
	\affiliation{J\"ulich Centre for Neutron Science JCNS at MLZ, Forschungszentrum J\"ulich, Lichtenbergstr. 1, D-85747 Garching, Germany}
	
	\author{Markus~Braden\,\orcidlink{0000-0002-9284-6585}}	
	\email{braden@ph2.uni-koeln.de}
	\affiliation{II. Physikalisches Institut, Universität zu Köln, Zülpicher Str. 77, D-50937 Köln, Germany}
	
	\author{Karin Schmalzl}
	\affiliation{J\"ulich Centre for Neutron Science (JCNS) at ILL, Forschungszentrum J\"ulich, F-38000 Grenoble, France}
	
	\author{Wolfgang Schmidt}
	\affiliation{J\"ulich Centre for Neutron Science (JCNS) at ILL, Forschungszentrum J\"ulich, F-38000 Grenoble, France}
	
	\author{Martin Meven}
	\affiliation{Institut f\"ur Kristallographie, RWTH Aachen, and J\"ulich Centre of Neutron Science at the Heinz Maier-Leibnitz Zentrum, Lichtenbergstr. 1, D-85747 Garching, Germany}
	
	\author{Erxi Feng}
	\affiliation{Institute of High Energy Physics, Chinese Academy of Sciences (CAS), Beĳing 100049, China}
	\affiliation{Spallation Neutron Source Science Center, Dongguan 523803, China}
	
	\author{Yinghao Zhu}
	\affiliation{J\"ulich Centre for Neutron Science JCNS at MLZ, Forschungszentrum J\"ulich, Lichtenbergstr. 1, D-85747 Garching, Germany}

	\author{Alexandre Bertin}
	\affiliation{II. Physikalisches Institut, Universität zu Köln, Zülpicher Str. 77, D-50937 Köln, Germany}
	
	\author{Paul Steffens}
	\affiliation{Institut Laue-Langevin, 71 avenue des Martyrs, CS 20156, 38042 Grenoble Cedex 9, France}
	
	\author{Yixi Su,\orcidlink{0000-0001-8434-1758}}
	\email{y.su@fz-juelich.de}
	\affiliation{J\"ulich Centre for Neutron Science JCNS at MLZ, Forschungszentrum J\"ulich, Lichtenbergstr. 1, D-85747 Garching, Germany}
	
	\date{\today}
	
	\begin{abstract}
		The layered honeycomb magnet \(\alpha\)-RuCl\(_3\) has attracted intense scrutiny as a prime candidate for realizing the Kitaev quantum spin liquid, yet a consensus on its microscopic Hamiltonian remains elusive due to the material's extreme sensitivity to structural details. Here, we report a comprehensive reexamination of the low-temperature crystallographic and magnetic structures of high-quality \(\alpha\)-RuCl\(_3\) single crystals using unpolarized and polarized neutron diffraction. We confirm a sharp, first-order structural phase transition to the rhombohedral \(R\bar{3}\) space group with a pronounced thermal hysteresis. Crucially, using both spherical and longitudinal neutron polarization analysis, we determine the 3D orientation of the ordered magnetic moment without the ambiguity typically arising from domain distributions. We find that the Ru\(^{3+}\) magnetic moments in the zigzag phase are tilted by \(15.7^\circ\) out of the hexagonal plane and, remarkably, exhibit an additional in-plane twist of \(-13.8^\circ\). This "tilted and twisted" geometry differentiates the ground state from the previously reported models based on unpolarized neutron diffraction or resonant elastic X-ray scattering (REXS) analysis. 
	\end{abstract}
	
	\maketitle
	
	The search for Kitaev quantum spin liquids (QSLs) in real materials has been largely driven by the exact solution of the Kitaev model, where bond-dependent Ising interactions on a honeycomb lattice lead to a ground state of fractionalized Majorana fermions and gauge fluxes~\cite{Kitaev2006,Jackeli2009,Chaloupka2015,Takayama2015Li,trebst2022}. The spin-orbit assisted Mott insulator \(\alpha\)-RuCl\(_3\) has emerged as the most promising candidate to realize this physics~\cite{Plumb2014,Johnson2015,Cao2016,Banerjee2017,KimAPL2022}. Extensive experimental efforts, including Raman spectroscopy, inelastic neutron scattering, and thermal transport measurements, have provided evidence for a proximate QSL state, characterized by a continuum of magnetic excitations and, most intriguingly, a potentially quantized thermal Hall effect in applied magnetic fields~\cite{Banerjee2016,Banerjee2017,KimAPL2022,Do2017,Ran2017,Kasahara2018,Balz2021,Czajka2021,Bruin2022,Braden2025}.
	
	However, \(\alpha\)-RuCl\(_3\) is not a pure Kitaev system. The Hamiltonian inevitably includes non-Kitaev terms such as the isotropic Heisenberg exchange \(J\) and symmetric off-diagonal \(\Gamma\) interactions~\cite{Chaloupka2010,Rau2014,Winter2016,Winter2017,Banerjee2017,Hermanns2018,Banerjee2018,Balz2019,Moller2025}. Similar physics has been explored in iridates such as Na$_2$IrO$_3$~\cite{Singh2010} and Li$_2$IrO$_3$~\cite{Takayama2015Li}, but RuCl$_3$ remains unique due to its stoichiometry and the availability of large single crystals. The interplay between the dominant ferromagnetic Kitaev term ($K$) and these perturbations determines the low-temperature ground state. In zero field, \(\alpha\)-RuCl\(_3\) undergoes a transition to long-range zigzag antiferromagnetic order below $T_N \approx 6\sim14$\,K~\cite{Johnson2015,Sears2015,KimKee2016,Sears2020,Zhang2023}. While the magnetic propagation vector is well established, the precise orientation of the ordered moments—which serves as a critical constraint for extracting the ratio of $K$, $J$, and $\Gamma$—remains controversial~\cite{Johnson2015,Cao2016,Maksimov2020,Sears2020,Liu2022,Ran2022,Kim2024,Park2024}.
	
	The difficulty in pinpointing the magnetic structure stems from the material's structural complexity. \(\alpha\)-RuCl\(_3\), as a member of the big Halides family~\cite{McGuire2017}, consists of weakly coupled van der Waals layers, making it prone to stacking faults and structural phase transitions~\cite{Cao2016,Mu2022,Zhang2024}. Early studies identified a monoclinic \(C2/m\) structure at room temperature, which was often assumed to persist or to define the symmetry down to low temperatures~\cite{Cao2016}. Based on this, magnetic models were constrained to high-symmetry directions allowed by the monoclinic mirror planes (i.e., moments in the monoclinic $ac$-plane). However, recent diffraction studies on high-quality crystals have revealed a transition to a rhombohedral \(R\bar{3}\) structure at low temperature~\cite{Mu2022,Sears2023,Park2024,Zhang2024,Kim2024}. Crucially, the $R\bar{3}$ stacking sequence explicitly breaks these single-layer mirror planes. This fundamentally changes the symmetry constraints on the moment direction, lifting the restriction that the spins must reside within the mirror plane and enabling an additional in-plane twist. This structural ambiguity has led to conflicting reports on the magnetic moment direction, with estimates for the out-of-plane canting angle ranging from $30^\circ$ to nearly $50^\circ$~\cite{Johnson2015,Cao2016,Sears2020,Kim2024,Park2024}. Muon spin experiments also suffered from the variation of transition temperatures and could only suggest a modulated zigzag spin structure~\cite{Yamauchi2018}.
	
	In this Letter, we resolve this controversy by combining unpolarized single-crystal neutron diffraction with advanced polarized neutron techniques~\cite{IN12data,Thales2023}. Exploiting the neutron's direct coupling to the total magnetic moment vector, we uniquely determine the magnetic ground state. We confirm the \(R\bar{3}\) low-temperature structure in high-quality single crystals. Most importantly, using Spherical Neutron Polarimetry (SNP)~\cite{Brown2005,Lelievre2005,IN12data} and longitudinal polarization analysis~\cite{Boehm2015,Braden2025,Thales2023}, we solve the full 3D orientation of the magnetic moments. We discover a ``tilted and twisted'' zigzag order with a substantial in-plane rotation component previously overlooked. This geometry provides strict new benchmarks for theoretical models of the anisotropic exchange interactions in \(\alpha\)-RuCl\(_3\).
	
	\begin{figure}[t!]
		\centering
		\includegraphics[width=0.9\columnwidth]{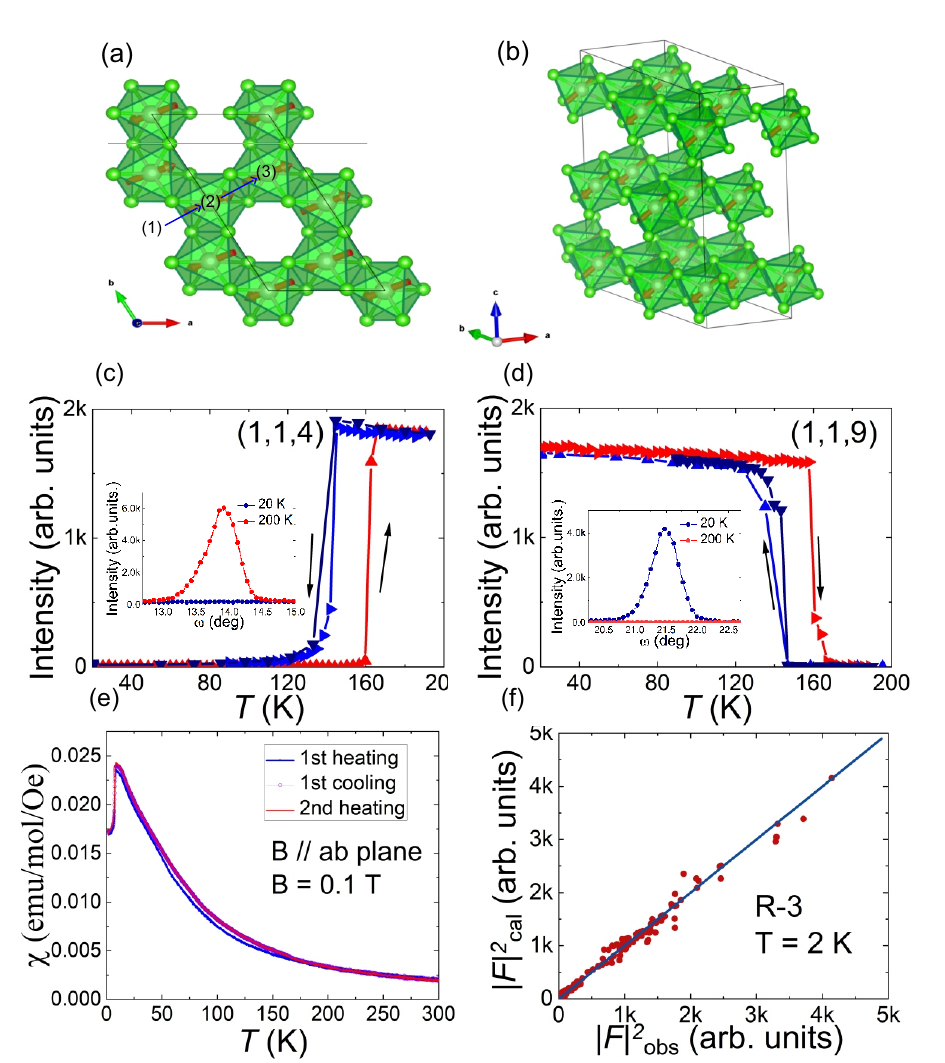}
		\caption{\label{fig:structure}
			Structures and phase transition of \arc.
			(a) View of a single honeycomb layer along the $c$-axis. The light-blue line indicates the mirror plane present in the single layer but broken in the \(R\bar{3}\) stacking. The labels (1--3) indicate the stacking vector positions. Crystal structure model was plotted by Vesta~\cite{Momma2011}.
			(b) 3D view of the unit cell.
			(c, d) Temperature dependence of the integrated intensities for Bragg peaks (1,1,4) and (1,1,9). The (1,1,4) reflection is allowed only in the monoclinic ($C2/m$) phase, while (1,1,9) is specific to the rhombohedral ($R\bar{3}$) phase. The clean hysteresis and lack of residual intensity at base temperature indicate a complete structural transition. Insets show rocking curves at 20\,K and 200\,K.
			(e) Magnetic susceptibility $\chi(T)$ with $B \parallel ab$, showing the structural hysteresis around 150\,K and the magnetic transition at low temperature.
			(f) Comparison of observed ($|F_{\mathrm{obs}}|^2$) and calculated ($|F_{\mathrm{cal}}|^2$) neutron diffraction structure factors for the $R\bar{3}$ model at 2\,K.
		}
	\end{figure}
	
	\textbf{\textit{Structural Phase Transition and $R\bar{3}$ Symmetry.---}} The presence of stacking faults is known to suppress the magnetic ordering temperature and broaden the transition. In "high-quality" samples, $T_N$ is typically around 7--8\,K, whereas samples with significant disorder order at temperatures as high as 14\,K~\cite{Cao2016,Kasahara2022thermal,Kim2024mb}. The crystals used in this study were grown by the chemical vapor transport method and selected for their sharp magnetic transitions ($T_N \approx 7.5$\,K) and clean diffraction profiles~\cite{Braden2025}.

	Single-crystal neutron diffraction was carried out on a $\sim$30 mg sample on the hot-neutron four-circle diffractometer HEiDi~\cite{Meven2015} at the Heinz Maier-Leibnitz Zentrum (MLZ). We investigated the structural phase transition by monitoring diagnostic nuclear Bragg peaks, indexed in the hexagonal setting ($a=b=5.973$\,\AA, $c=16.93$\,\AA, $\alpha=\beta=90^\circ$, $\gamma=120^\circ$). In the following we always refer to this hexagonal lattice in obverse setting and all scattering vectors are given in the corresponding reduced lattice vectors. To verify the completeness of the structural transition, we utilized the $(1,1,4)$ and $(1,1,9)$ reflections as mutually exclusive markers. In the $R\bar{3}$ space group, allowed reflections must satisfy the rhombohedral centering condition $-h+k+l=3n$ for the obverse setting (or $h-k+l=3n$ for the reverse twin). Thus, $(1,1,4)$ is strictly forbidden in the pure $R\bar{3}$ ground state but allowed in the high-temperature $C2/m$ phase. Conversely, $(1,1,9)$ satisfies the $R\bar{3}$ rule but is forbidden in $C2/m$. The detailed index mappings between the $R\bar{3}$ and $C2/m$ phases are discussed in the Supplementary Material~\cite{supmat}. As shown in Fig.~\ref{fig:structure}(c,d), the $(1,1,4)$ reflection vanishes completely upon cooling below 120\,K, while the $(1,1,9)$ reflection concurrently appears with high intensity. The transition is first-order, evidenced by a broad thermal hysteresis. This simultaneous exchange of peak intensity unambiguously confirms a complete transformation into the $R\bar{3}$ structure, with no residual high-temperature phase detected at 2\,K.
	
	We collected a dataset of 601 reflections at 2\,K to refine the structure. The refinement (Fig.~\ref{fig:structure}(f), Table~\ref{tab:r-3_struture}) yields a good fit with the \(R\bar{3}\) space group, accounting for obverse and reverse rhombohedral twins using the Jana2006 software~\cite{Petricek2016}. In the \(R\bar{3}\) stacking, neighboring layers are shifted parallel to a Ru-Ru bond (see Fig.~\ref{fig:structure}(a)). This is physically distinct from the \(C2/m\) stacking, where shifts occur perpendicular to a bond. The \(R\bar{3}\) symmetry breaks the mirror planes of the individual honeycomb layers.
	
	This symmetry breaking has a subtle but significant effect on the local crystal field. In the \(R\bar{3}\) structure, the Ru atoms sit on a threefold axis, but the upper and lower triangular faces of the Cl$_6$ octahedra are not equivalent. The Cl$_3$ triangle facing a Ru in the neighboring RuCl$_3$ layer is expanded (Cl-Cl distance 3.396\,\AA) compared to the triangle facing the void in the opposite neighboring layer (3.360\,\AA). This $\sim 1\%$ difference in bond lengths introduces a trigonal distortion that is absent in the ideal single-layer limit and distinct from the monoclinic distortion. Given the sensitivity of the Kitaev and $\Gamma$ terms to bond angles and distances~\cite{Winter2016}, this structural nuance is critical for accurate modeling of the spin Hamiltonian.  In spacegroup  \(R\bar{3}\) the Ru site is not fixed to $z$=1/3, but we do not find a significant deviation, see Table I. Nevertheless, the significant deviation for the Cl sites excludes a higher symmetry.
	
	\begin{table}[htbp]
		\caption{\label{tab:r-3_struture}
			Refinement results with single-crystal neutron diffraction data for \(\alpha\)-RuCl\(_3\) at \(T=2\,\mathrm{K}\) for the \(R\bar{3}\) structural model in obverse hexagonal setting. Atomic coordinates, site occupancy, and atomic displacement parameters (ADPs, given in \AA$^{-1}$) are listed.
		}
		\small
		\begin{center}
			\begin{tabular}{cccccccc}
				\hline\hline
				& Atom & Site & \(x\) & \(y\) & \(z\) & Occ. & \\
				\hline
				& Ru & \(6c\) & 0 & 0 & 0.3332(3) & 1 & \\
				& Cl & \(18f\) & 0.3203(4) & 0.3356(4) & 0.4114(8) & 1 & \\
				\hline
				& \multicolumn{7}{l}{{Harmonic ADPs given in \AA$^2$:}} \\
				\hline
				&  & \(U_{11}\) & \(U_{22}\) & \(U_{33}\) &  &  & \\
				\hline
				& Ru & 0.0033(10) & 0.0033(10) & 0.0107(14) &  &  & \\
				& Cl & 0.0063(9)  & 0.0040(8)  & 0.0069(5)  &  &  & \\
				\hline
				&  & \(U_{12}\) & \(U_{13}\) & \(U_{23}\) &  &  &  \\
				\hline
				& Ru & 0.0016(5) & 0         & 0         &  &  &  \\
				& Cl & 0.0029(7) & 0.0021(5) & -0.0007(4) &  &  &  \\
				\hline
				& \multicolumn{7}{l}{Twin population: 0.44(4)} \\
				& \multicolumn{7}{l}{\(\displaystyle a = b = 5.973\,\text{\AA}, \quad c = 16.93\,\text{\AA}\)} \\
				\hline
				& \multicolumn{7}{l}{\(R(\mathrm{obs}) = 7.45\%\)} \\
				\hline\hline
			\end{tabular}
		\end{center}
	\end{table}
	
	\textbf{\textit{Solving the Magnetic Structure with Polarized Neutrons.---}} The zigzag magnetic order in the \(R\bar{3}\) phase is described by the propagation vector $\bm{k}=(0, 0.5, 1)$ in obverse hexagonal notation. The magnetic ordering breaks the threefold rotation symmetry of spacegroup  \(R\bar{3}\), reducing the magnetic space group to triclinic $P_S\bar{1}$~\cite{PerezMato2015,isodistort1,isodistort2}. In this low-symmetry group, the magnetic moment $\bm{M}$ is unconstrained and can have non-zero components along the three orthogonal directions $a$, $b^\star$ and $c$. Note that this differs from the magnetic symmetry in the monoclinic structure, which does not permit a component along the monoclinic $b_m$ direction.
	
    Standard unpolarized neutron diffraction struggles to uniquely solve this structure because of the inherent coupling between the magnetic moment direction and the domain populations. The structural transition creates two types of twin domains (obverse and reverse), and the magnetic ordering creates three possible $120^\circ$ domains~\cite{PerezMato2015,isodistort1,isodistort2} in each of the structural domains, as it is shown in the Fig.~\ref{fig:mag_structure}(a) for obverse domains, see also the Supplemental Material~\cite{supmat}. Neglecting the time-inversion or AFM 180-degrees domains, there are thus 6 different magnetic domains in the magnetic phase of \arc , which also differs to the situation in the $C2/m$ symmetry, where zigzag AFM stacking is tight to the monoclinic axis \cite{C2m-note}.

   To disentangle the magnetic moment direction from the domain population and scale factors, we employed SNP using the Cryopad device on the IN12 spectrometer at the ILL~\cite{Schmalzl2016in12,Lelievre2005}. Since we probe a single magnetic domain in reciprocal space, the measured polarization matrix elements $P_{ij}$, which represent ratios of spin-dependent cross-sections, are independent of the domain volume-fraction and of the absolute moment size. This allows us to uniquely determine the moment orientation $(\theta, \beta)$ without relying on assumptions about the domain distribution. The out-of-plane tilt $\theta$ is measured from the $ab$ plane, and the in-plane twist $\beta$ is defined such that $\beta = 0^\circ$ confines the magnetic moment entirely within this $ac$ plane.
	
	For a collinear magnetic structure, the diagonal elements of the polarization matrix depend on squares of the magnetic components perpendicular to $\bm{Q}$ ($M_{\perp y}^2$ and $M_{\perp z}^2$). Crucially, the off-diagonal elements (e.g., $P_{yz}, P_{zy}$) depend on the \textit{product} $M_{\perp y} M_{\perp z}$~\cite{Brown2001,Qureshi2020}. This phase sensitivity allows SNP to determine the relative orientation of the moment components precisely.
	
\begin{figure}[htbp]
		\centering
		\includegraphics[width=\columnwidth]{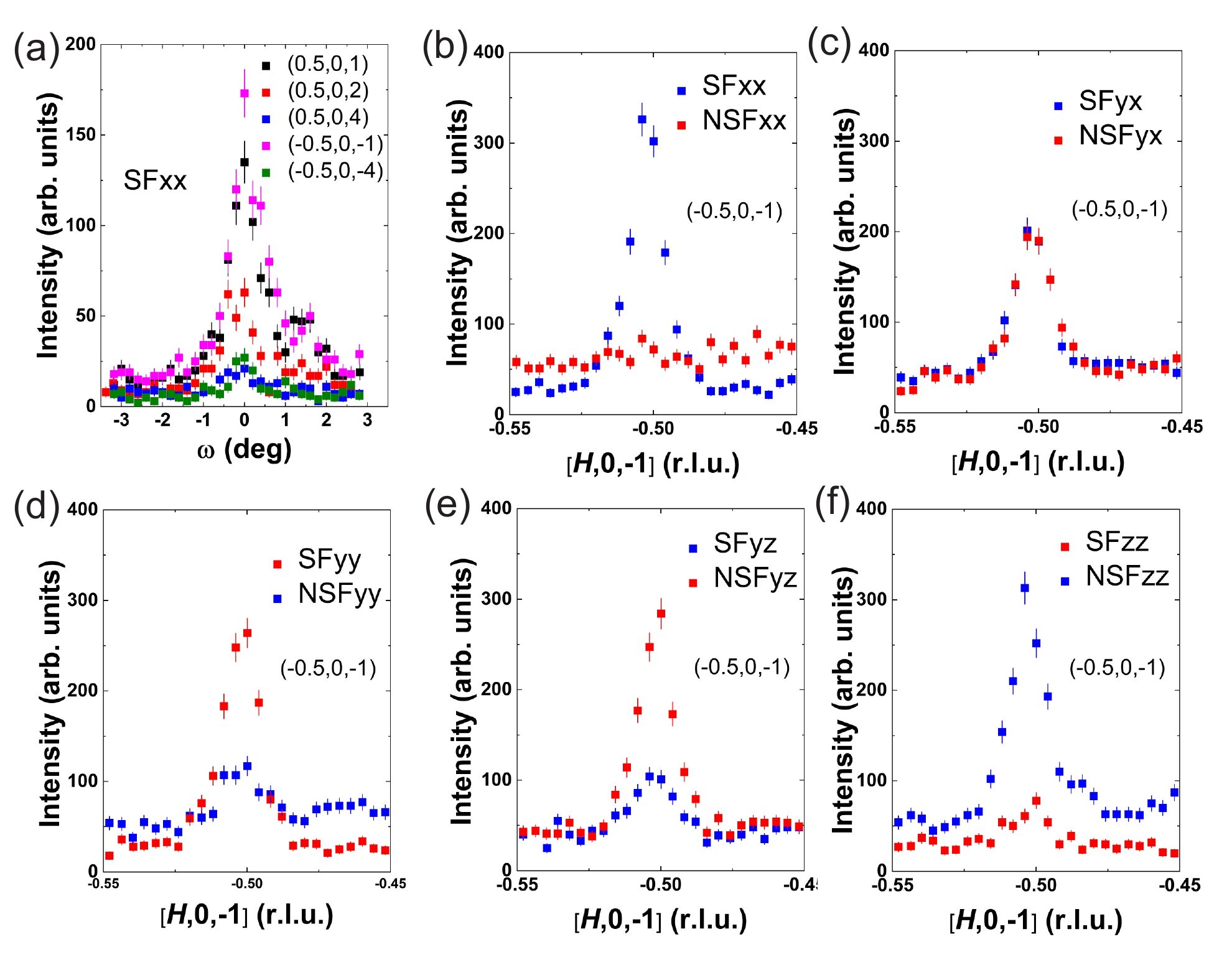}
		\caption{\label{fig:snp}
			Spherical neutron polarimetry (SNP) analysis at $T=1.5$\,K. $\text{SF}_{ij}$ and $\text{NSF}_{ij}$ denote the spin-flip and non-spin-flip scattering cross-sections, respectively, where the subscripts $i, j \in \{x, y, z\}$ indicate the polarization directions of the incident and scattered neutrons. 
			(a) Rocking curves ($\omega$-scans) of key magnetic reflections measured in the $\text{SF}_{xx}$ channel. 
			(b)-(f) $Q$ scans across the $(-0.5, 0, -1)$ reflection, resolved into various SF and NSF polarization channels. The local coordinate system is defined with $x \parallel \mathbf{Q}$ and $z$ perpendicular to the scattering plane (vertical). The absence of peak intensity in the $\text{NSF}_{xx}$ channel (b) confirms the purely magnetic origin of this reflection. Furthermore, the substantial intensity in the off-diagonal channels, such as $\text{SF}_{yz}$ (e) and $\text{SF}_{yx}$ (c), directly probes the cross-terms of the magnetic interaction vector. This provides the tight constraints necessary to uniquely determine the tilted and twisted moment direction.
		}
	\end{figure}

\begin{figure}[htbp]
		\centering
		\includegraphics[width=0.9\columnwidth]{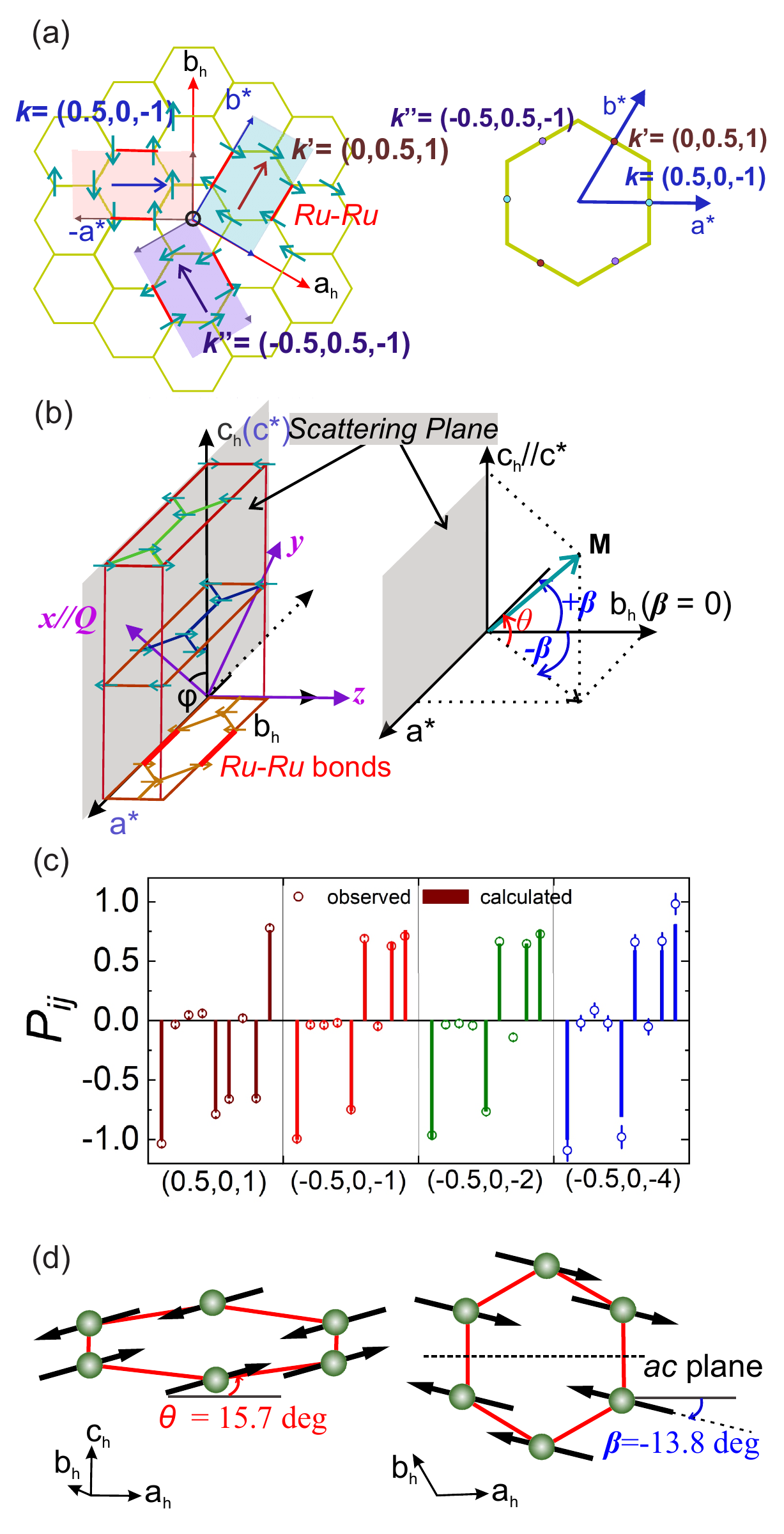}
		\caption{\label{fig:mag_structure}
			Determination of the magnetic moment direction by the SNP method. 
				(a) Three zigzag magnetic configuration domains in the obverse hexagonal structure. Their magnetic configurations and propagation directions are marked. The right panel shows their corresponding ordered positions in reciprocal space. In our SNP scattering geometry, we exclusively access one domain, as the other two domains point out of the scattering plane. 
				(b) Scattering geometry and definition of the $xyz$ coordinate system used for the SNP experiment. The moment direction is defined by the tilt angle $\theta$ (out of the $ab$ plane) and the twist angle $\beta$ (in-plane rotation).
				(c) Comparison of observed (open circles) and calculated (solid bars) polarization matrix elements $P_{ij}$. The sequence of $P_{ij}$ for each reflection is plotted as $P_{xx}$, $P_{xy}$, $P_{xz}$, $P_{yx}$, $P_{yy}$, $P_{yz}$, $P_{zx}$, $P_{zy}$, $P_{zz}$. 
				(d) The determined zigzag magnetic structure. The best fit yields moment orientation angles of $\theta = 15.7(1.1)^\circ$ and $\beta = -13.8(1.5)^\circ$, indicating a ``tilted (out-of-plane) and twisted (in-plane)'' geometry. 
				 A twist of $\beta = 0^\circ$ confines the moment to the hexagonal $ac$ plane when the antiparallel bond of the zigzag order points along $b^\star$,
				and negative $\beta$ value indicates a clockwise rotation away from the hexagonal $ac$ plane.
		}
	\end{figure}

	We measured the full polarization matrix for the magnetic reflections $(0.5, 0, 1)$, $(-0.5, 0, -1)$, $(0.5, 0, 2)$, and $(0.5, 0, 4)$. As shown in the representative scans in Fig.~\ref{fig:snp}, we observed significant intensities in the off-diagonal channels. We parameterized the magnetic moment direction using two angles: $\theta$, the canting angle out of the honeycomb ($ab$) plane, and $\beta$, the in-plane rotation angle (where $\beta=0^\circ$ corresponds to moments perpendicular to the Ru-Ru bond), as shown in Fig.~\ref{fig:mag_structure}(a) and (b). We performed a least-squares fit to the experimental polarization matrices to determine these angles. The detailed SNP data analysis methods are provided in the Supplementary Material~\cite{supmat}. The best fit, shown in Fig.~\ref{fig:mag_structure}(c), yields $\theta = 15.7(1.1)^\circ$ and $\beta = -13.8(1.5)^\circ$ as shown in Fig.~\ref{fig:mag_structure}(e).
	
	This result is striking for two reasons. First, the out-of-plane tilt angle $\theta \approx 15.7^\circ$ is significantly smaller than the values of $30^\circ-50^\circ$ reported in earlier neutron and X-ray studies~\cite{Cao2016,Park2024}. Second, and more remarkably, we find a non-zero in-plane twist of $\beta \approx -13.8^\circ$, leading us to characterize this magnetic structure as ``tilted and twisted.'' Most previous theoretical and experimental works have assumed $\beta=0^\circ$ based on the symmetry constraints of the monoclinic phase~\cite{Sears2020,Singh2010}. Our data strongly exclude this $\beta=0^\circ$ scenario: forcing $\beta=0^\circ$ in our refinement yields a best-fit $\theta$ value of $24.5^\circ$, but results in a sizable increase in the goodness-of-fit to $\chi^2 = 210$, compared to $\chi^2 = 117$ when $\beta$ is refined freely (as shown in Fig. S2 of the Supplementary Material~\cite{supmat}). This indicates that the in-plane twist is an intrinsic feature of the ground state.
	
	\textbf{\textit{Verification via Longitudinal Polarization.---}} To validate this ``tilted and twisted'' model, we performed an independent experiment using longitudinal polarization analysis on the Thales spectrometer~\cite{Boehm2015,Thales2023,Braden2025} with a larger ($\sim 700$\,mg) single crystal, which was carefully characterized to ensure a negligible density of stacking faults~\cite{Braden2025}. In this setup, we mainly measured the diagonal spin-flip cross-sections $\text{SF}_{yy}$ and $\text{SF}_{zz}$ for a series of magnetic reflections $(0, 0.5, L)$ with $L=1, 2, 4, 5$.
	\begin{figure}[htbp]
		\centering
		\includegraphics[width=0.95\columnwidth]{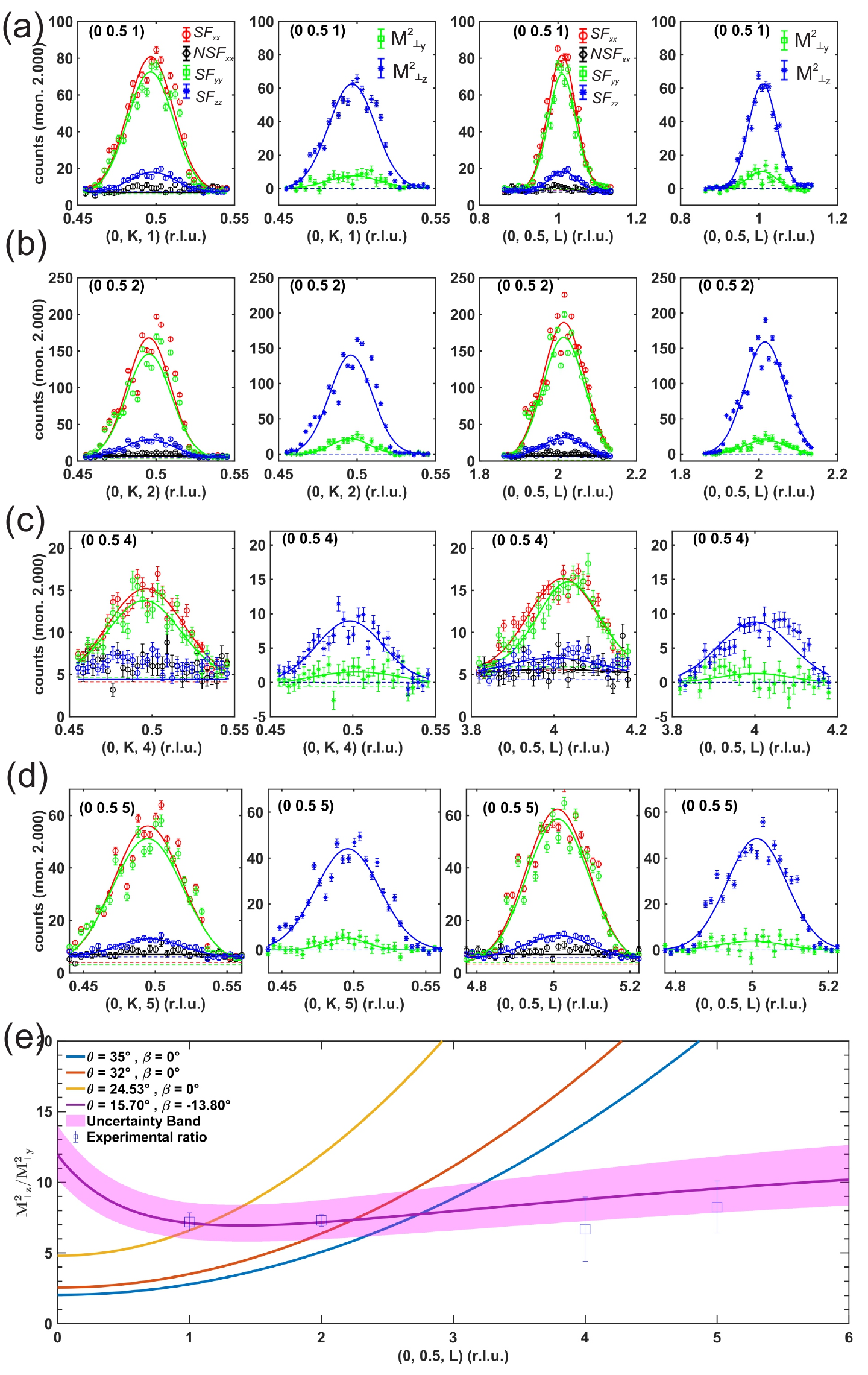}
		\caption{\label{fig:xyz_scans}
			Differentiation of magnetic structure models via XYZ-polarization analysis.
			(a)--(d) $Q$-scans along $K$ and $L$ directions for magnetic reflections $(0, 0.5, L)$ with $L=1, 2, 4, 5$. Data were collected in the $x, y, z$ spin-flip ($\text{SF}_{xx}$, $\text{SF}_{yy}$, $\text{SF}_{zz}$) and $x$ non-spin-flip ($\text{NSF}_{xx}$) channels. For each scan direction, the isolated magnetic intensity components are obtained via the subtractions $M_{\perp z}^2 \sim \text{SF}_{xx} - \text{SF}_{zz}$ and $M_{\perp y}^2 \sim \text{SF}_{xx} - \text{SF}_{yy}$.
			(e) The ratio of magnetic intensity components $M_{\perp z}^2 / M_{\perp y}^2$ as a function of $L$. Symbols represent experimental values derived from the fits in (a)--(d). The curves represent calculated ratios for different magnetic moment orientations: blue ($\theta = 35^\circ, \beta = 0^\circ$)~\cite{Cao2016}, orange ($\theta = 32^\circ, \beta = 0^\circ$)~\cite{Sears2020}, and yellow ($\theta = 24.53^\circ, \beta = 0^\circ$) obtained by forcing $\beta = 0^\circ$ in the SNP fit. The solid magenta line indicates the best-fit structure determined from the full SNP analysis ($\theta \approx 15.7^\circ, \beta \approx -13.8^\circ$), with the shaded region representing the uncertainty band. The data clearly exclude the $\beta=0^\circ$ models and confirm the twisted moment structure.
		}
	\end{figure}
	
	In longitudinal analysis, the neutron spin follows the guide field. The selection rules dictate that $SF_{yy}$ intensity arises purely from the magnetic component $M_{\perp z}$ that always points along $a$, while $SF_{zz}$ intensity arises from $M_{\perp y}$~\cite{Jenni2022}. Therefore, the ratio $SF_{yy}/SF_{zz} = M_{\perp z}^2 / M_{\perp y}^2$ provides a direct probe of the moment anisotropy that is independent of domain volume fractions, form factors, and instrumental absolute scaling.
	
	Figure~\ref{fig:xyz_scans}(e) displays this ratio as a function of $L$. For a magnetic moment confined to the $ac$-plane ($\beta=0$), geometric projection effects cause this ratio to increase sharply at large $L$ values (Fig.~\ref{fig:xyz_scans}(e)). Our experimental data, however, show a much flatter dependence on $L$. This behavior is excellently described by the model derived from our SNP analysis (solid magenta line), which incorporates a finite twist angle $\beta \approx -13.8^\circ$. The quantitative agreement between these two distinct polarized neutron techniques provides robust evidence that the magnetic moments in $\alpha$-RuCl$_3$ deviate from high-symmetry directions. Furthermore, the consistency of results between the small crystal ($\sim$100\,mg) measured at IN12 and the large crystal ($\sim$700\,mg) measured at Thales demonstrates the robustness of this magnetic structure across different samples.
		
	\textbf{\textit{Discussion.---}} The two polarized neutron experiments perfectly agree, ruling out the previous models. A detailed comparison of the simulated REXS azimuthal scan for the model obtained in this study and for that obtained via REXS as well as the discussions on the found discrepancy and possible causes are given in the Supplementary Material~\cite{supmat}.
	
	Physically, the exact orientation of the ordered moment heavily constrains the microscopic interaction parameters~\cite{Maksimov2020}. While our out-of-plane canting sign agrees with REXS~\cite{Sears2020,Kim2024}---supporting a ferromagnetic Kitaev interaction---our absolute tilt is smaller and accompanied by an in-plane twist. This discrepancy arises because the techniques sense different vectors. Neutrons measure the total magnetic moment (spin plus orbital). Due to strong $g$-factor anisotropy ($g_{a,b} = 2.53$, $g_c = 1.56$~\cite{Suzuki2021}), this total moment is not parallel to the pseudospin $j$ vector~\cite{Braden2025,Chaloupka2016}. Correcting our measured $\theta=15.7^\circ$ tilt for the $g$-factor anisotropy yields a $24.5^\circ$ pseudospin canting. The  REXS method also does not directly sense the pseudospin but a sum of spin and orbital contributions with distinct weighting, which can explain the even larger canting  value obtained in these experiments.
	
	Finally, the finite twist ($\beta \approx -13.8^\circ$) highlights the crucial role of magnetoelastic coupling. Recent thermal expansion and strain studies~\cite{Li2025} reveal a spontaneous in-plane structural anisotropy below $T_N$, indicating that the system lowers its effective interaction symmetry by splitting the anisotropic interaction along the inequivalent nearest- and further nearest bonds. Spontaneous structural distortions locally alter the balance of bond-dependent anisotropic exchange interactions (e.g., Kitaev $K$ and off-diagonal $\Gamma$ terms) which also allow for fine tuning of magnetic exctations~\cite{Braden2025}. Consequently, $\beta$ is energetically pinned relative to the selected zigzag direction. The twist is therefore not an arbitrary variable, but a strict energetic minimum dictated by the magnetoelastically modulated spin Hamiltonian.
	
	\textbf{\textit{Summary.---}} Using spherical and longitudinal neutron polarization analysis, we determined the direction of the antiferromagnetic moments of high-quality $\alpha$-RuCl$_3$ single crystals in the rhombohedral $R\bar{3}$ phase. By directly measuring off-diagonal polarization matrix elements and by analyzing intensity ratios in longitudinal neutron polarization, we unambiguously resolve the 3D orientation of the zigzag antiferromagnetic order, revealing an out-of-plane canting of $\theta = 15.7^\circ$ and an in-plane twist of $\beta = -13.8^\circ$. We emphasize that this twist is an intrinsic feature permitted by the $R\bar{3}$ space group, which lacks previously assumed mirror planes. Furthermore, correcting our measured macroscopic tilt for the known $g$-factor anisotropy yields a pseudospin canting of $24.5^\circ$. Accurately resolving this bulk 3D moment direction provides crucial constraints for determining the effective spin Hamiltonian in $\alpha$-RuCl$_3$.
	
	\textbf{\textit{Acknowledgments.---}} X. Wang is supported by the National Natural Science Foundation of China (Grant No. 12505350). F. Zhu is supported by the National Key Research and Development Program of China (Grant No. 2022YFA1405700).  E. Feng is supported by the National Natural Science Foundation of China (Grant No. 12375298). Y. Zhu acknowledges the postdoctoral funding from the European Union’s Horizon 2020 research and innovation program under the Marie Skłodowska-Curie Grant Agreement No. 101034266. We also acknowledge support by the Deutsche Forschungsgemeinschaft (DFG, German Research Foundation) - Project number 277146847 - CRC 1238, project B04. We thank Thomas Br\"uckel and Georg Roth for valuable discussions. Unpolarized neutron diffraction data were taken on the single crystal diffractometer HEiDi, operated jointly by RWTH Aachen University and Forschungszentrum Jülich GmbH within the JARA alliance. We thank Markos Skoulatos and Bastian Veltel for their assistance in the DC magnetisation measurements in the Physics Lab. at the Heinz Maier-Leibnitz Zentrum (MLZ), Garching.

	\bibliography{RuCl3_cpl_combined.bib}
	
	\clearpage
	

\section*{Supplementary material}

# Supplementary Material for:
# Tilted and Twisted Magnetic Moments in the Kitaev Magnet $\alpha$-RuCl$_3$

Xiao Wang,[1,2,3] Fengfeng Zhu,[4,3] Markus Braden,[5,*] Karin Schmalzl,[6] Wolfgang Schmidt,[6] Martin Meven,[7] Erxi Feng,[1,2] Yinghao Zhu,[3] Alexandre Bertin,[5] Paul Steffens,[8] and Yixi Su[3,†]

[1] Institute of High Energy Physics, Chinese Academy of Sciences (CAS), Beijing 100049, China
[2] Spallation Neutron Source Science Center, Dongguan 523803, China
[3] Jülich Centre for Neutron Science JCNS at MLZ,
Forschungszentrum Jülich, Lichtenbergstr. 1, D-85747 Garching, Germany
[4] State Key Laboratory of Materials for Integrated Circuits, 2020 X-Lab,
Shanghai Institute of Microsystem and Information Technology,
Chinese Academy of Sciences, Shanghai 200050, China
[5] II. Physikalisches Institut, Universität zu Köln, Zülpicher Str. 77, D-50937 Köln, Germany
[6] Jülich Centre for Neutron Science (JCNS) at ILL,
Forschungszentrum Jülich, F-38000 Grenoble, France
[7] Institut für Kristallographie, RWTH Aachen, and Jülich Centre of Neutron Science
at the Heinz Maier-Leibnitz Zentrum, Lichtenbergstr. 1, D-85747 Garching, Germany
[8] Institut Laue-Langevin, 71 avenue des Martyrs, CS 20156, 38042 Grenoble Cedex 9, France
(Dated: March 30, 2026)

## SAMPLE PREPARATION AND NEUTRON SCATTERING

High-quality single crystals of $\alpha$-RuCl$_3$ were grown by the chemical vapor transport technique in a temperature gradient, utilizing a self-transport (sublimation) mechanism without the use of an external transport agent. High-purity anhydrous $\alpha$-RuCl$_3$ powder (Furuya Metal, Japan) was sealed in an evacuated quartz ampoule and placed in a two-zone furnace. Driven by the difference in saturation vapor pressure, the material was transported from the source zone at 780 °C to the sink zone at 650 °C over a duration of five days.

Single-crystal neutron diffraction and structural investigations were carried out on a ∼30 mg single crystal sample on the hot-neutron four-circle diffractometer HEiDi [1] at the Heinz Maier-Leibnitz Zentrum (MLZ) using a wavelength of 0.81 Å. Spherical Neutron Polarimetry (SNP) measurements were performed on the cold triple-axis spectrometer IN12 [2] at the Institut Laue-Langevin (ILL), Grenoble, with a wave vector $k_f = 2\,\text{Å}^{-1}$. For this study a ∼100 mg single crystal of $\alpha$-RuCl$_3$ was mounted in the $(h,0,l)$ scattering plane (with respect to the hexagonal lattice of space-group $R\bar{3}$). The zero-field polarimeter *Cryopad* was installed to collect polarized data in 9 different polarization channels. Diffraction measurements with longitudinal neutron-polarization were conducted on the cold triple-axis spectrometer Thales [3] at ILL using another single crystal weighing approximately 700 mg [4]. Focusing Heusler crystals served as both the polarizing monochromator and analyzer, and a Helmholtz coil arrangement allowed precise adjustment of the neutron polarization at the sample position. To suppress higher-order contamination, a velocity selector and a beryllium filter were employed, and most measurements were performed at a fixed final wave vector of $k_f = 1.50\,\text{Å}^{-1}$. Flipping ratios were measured on the nuclear Bragg peaks for both polarized neutron experiments for calibration of the incident neutron beam. On Thales these amount to 27 and 20 at the nuclear Bragg peaks (0,0,3) and (0,0,6), respectively. The slightly lower value at the larger scattering arises from cross talks between the guide field and is less relevant for the magnetic measurements. On IN12 we obtained a flipping ratio of 24.9.

## REFINEMENT OF C2/M STRUCTURE

For the refinement of the monoclinic model, we first indexed the reflections measured in the hexagonal setting into the monoclinic setting [5]. For the $C2/m$ structure model, we consider three domains related by a pseudo threefold symmetry operator along the $c^*$ direction. The unit cells for the three monoclinic domains are constructed using the following cell transformations:

- Domain 1: $\mathbf{a}_1 = \mathbf{a}_h$; $\mathbf{b}_1 = \mathbf{a}_h + 2\mathbf{b}_h$; $\mathbf{c}_1 = (-\mathbf{a}_h + \mathbf{c}_h)/3$.

- Domain 2: $\mathbf{a}_2 = \mathbf{b}_h$; $\mathbf{b}_2 = 2\mathbf{a}_h - \mathbf{b}_h$; $\mathbf{c}_2 = (-\mathbf{b}_h + \mathbf{c}_h)/3$.

- Domain 3: $\mathbf{a}_3 = -\mathbf{a}_h - \mathbf{b}_h$; $\mathbf{b}_3 = \mathbf{a}_h - \mathbf{b}_h$; $\mathbf{c}_3 = (\mathbf{a}_h + \mathbf{b}_h + \mathbf{c}_h)/3$.

Here, $\mathbf{a}_i$, $\mathbf{b}_i$, and $\mathbf{c}_i$ ($i = 1,2,3$) are the unit cell axes for the three monoclinic domains, and $\mathbf{a}_h$, $\mathbf{b}_h$, $\mathbf{c}_h$ are the unit cell axes for the rhombohedral structure in a hexagonal setting. Not all reflections could be converted to monoclinic cells; notably, 122 reflections — many of which are strong — could not be indexed in any of the three monoclinic domains.

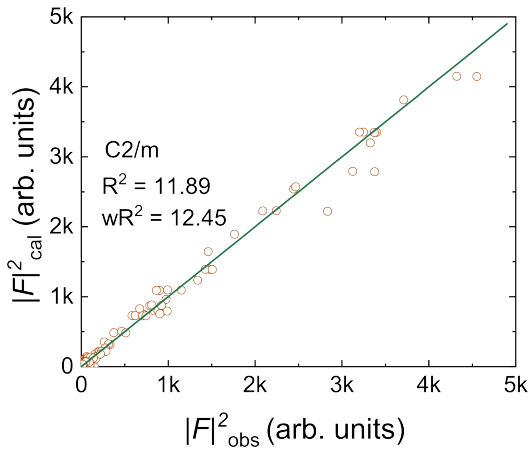

FIG. 1. Comparison of observed ($|F_{\mathrm{obs}}|^2$) and calculated ($|F_{\mathrm{cal}}|^2$) structure factors for the $C2/m$ structure model at $T = 2\,\mathrm{K}$. Single-crystal neutron diffraction data were collected, and 122 reflections that could not be indexed in any of the monoclinic domains were excluded from the refinement. The resulting reliability factors are $R = 11.89\%$.

The refinement results are shown in Fig. 1 and Table I. The $C2/m$ refinement yields results that are inferior to those obtained with the $R\bar{3}$ structure model and the exclusion of 122 reflections — especially those with strong intensities — strongly indicates that the low-temperature structure is not monoclinic.

TABLE I. Refinement results for $\alpha$-RuCl$_3$ from single-crystal neutron diffraction data at $T = 2\,\mathrm{K}$ using the $C2/m$ structural model.

| Atom | Site | $x$ | $y$ | $z$ | Occ. | $U_{iso}$ |
|---|---|---|---|---|---|---|
| Ru | $4g$ | 0 | 0.3343(5) | 0 | 1 | 0.0057(9) |
| Cl1 | $8j$ | 0.7447(10) | 0.1748(3) | 0.7643(4) | 1 | 0.0062(7) |
| Cl2 | $4i$ | 0.7278(11) | 0 | 0.2348(5) | 1 | 0.0062(7) |
| Domain populations: 0.319(5); 0.302(2) | | | | | | |
| $a = 5.973\,\text{Å}$, $b = 10.346\,\text{Å}$, $c = 5.983\,\text{Å}$, $\beta = 109.43°$ | | | | | | |
| $R(\mathrm{obs}) = 11.89\%$ | | | | | | |

## MAGNETIC DOMAINS

The triple-layer stacking of the antiferromagnetic zigzag structure is described by the propagation vector (0,0.5,1) in the obverse hexagonal setting, which leads to the magnetic space group $P_S\bar{1}$ [6–8]. There is only a single magnetic site with three independent components, and moments at all other Ru sites are either parallel or antiparallel, i.e. the magnetic structure is collinear. This contrasts to the analysis for the triple-stacking in space group $C2/m$, which allows more complex structures. The (0,0.5,1) propagation vector in $R\bar{3}$ signifies that the extra translation in obverse setting yields a $\pi$ phase shift, or antiparallel moments. Note that three times this translation corresponding to $(2a,b,c)$ implies also a $\pi$ phase shift, but this signifies magnetic order in phase with that in the initial layer due to the $b$ translation. Magnetic Bragg peaks occur thus at $(0,0.5,1)+(h,k,l)$ with the indices satisfying the obverse selection rule given above, such as $(0,0.5,1\pm 3n)$ with integer $n$. The reflections $(0,0.5,-1\pm 3n)$ do not belong to this series but corresponds to the reverse rhombohedral twin with the reverse translation and selection rule. The propagation vectors cannot be mixed, for example (0,0.5,-1) in the obverse domain yields a different magnetic structure with three independent magnetic moments. For completeness we add that the in-phase coupling for the obverse translation yields a sixfold stacking of the zigzag layers and can be described by a propagation vector (0,0.5,-0.5) which is realized in the intermediate magnetic phase at high magnetic fields. Besides antiphase domains, the zigzag magnetic ordering in $R\bar{3}$ results in three domains corresponding to the three possible directions of the zigzag stacking for the obverse and reverse rhombohedral structural domains [7, 8]. In particular there are no additional domains associated with the sign of the twisting which results in different magnetic structures.

## SPHERICAL POLARIZATION ANALYSIS

### Reducing the Polarization Matrix

In polarized neutron scattering, the scattered neutron spin polarization $\mathbf{P}_f$ is related to the incident neutron spin $\mathbf{P}_i$ via the tensor equation $\mathbf{P}_f = \mathcal{P}\,\mathbf{P}_i + \mathbf{P}'$, where $\mathcal{P}$ is a matrix describing the rotation of the polarization and $\mathbf{P}'$ is the polarization generated during scattering [9, 10]. In an experiment one specifies the directions of incoming ($i$) and outgoing ($f$) polarization directions and measures the SF$_{fi}$ and NSF$_{fi}$ intensities in nine distinct channels. Thereby one constructs the polarization matrix, which contains information about both the rotation and the generated polarization but which does not directly correspond to the transformation matrix $\mathcal{P}$ given above. Each element of the polarization matrix is determined experimentally as $P_{fi} = \frac{NSF_{fi} - SF_{fi}}{NSF_{fi} + SF_{fi}}$.

Theoretically, the polarization matrix can be expressed as follows. To handle the complex expression, we use a wide text format:

$$\begin{pmatrix} \frac{p_0 \tilde{I} - J_{yz}}{I} & \frac{p_0 J_{nz} - J_{yz}}{I} & \frac{p_0 J_{ny} - J_{yz}}{I} \\ \frac{-p_0 J_{nz} + R_{ny}}{I} & \frac{p_0 (\tilde{I} + 2M_{\perp y}^2) + R_{ny}}{I} & \frac{p_0 R_{yz} + R_{ny}}{I} \\ \frac{-p_0 J_{ny} + R_{nz}}{I} & \frac{p_0 R_{zy} + R_{nz}}{I} & \frac{p_0 (\tilde{I} + 2M_{\perp z}^2) + R_{nz}}{I} \end{pmatrix} \quad (1)$$

with

$$N^2 = NN^*, \quad \tilde{I} = N^2 - M_{\perp y}^2 - M_{\perp z}^2$$
$$R_{ni} = 2\,\mathrm{Re}\!\left(N\,M_{\perp i}^*\right), \quad R_{ij} = 2\,\mathrm{Re}\!\left(M_{\perp i}\,M_{\perp j}^*\right),$$
$$J_{ni} = 2\,\mathrm{Im}\!\left(N\,M_{\perp i}^*\right), \quad J_{ij} = 2\,\mathrm{Im}\!\left(M_{\perp i}\,M_{\perp j}^*\right),$$
$$I = N^2 + M_{\perp y}^2 + M_{\perp z}^2 + p_x J_{yz} + p_y R_{ny} + p_z R_{nz}.$$

Here, $N$ is the nuclear structure factor, and the magnetic interaction vector, $\mathbf{M} = \hat{\mathbf{Q}} \times (\mathbf{M} \times \hat{\mathbf{Q}})$ represents the component of the ordered moment perpendicular to $\mathbf{Q}$. The quantities $M_{\perp y}$ and $M_{\perp z}$ are the magnetic components along the $y$ and $z$ directions, respectively, and $p_0$ denotes the incident beam polarization with coefficients $(p_x, p_y, p_z)$. Note that the intensity $I$ depends on the initial polarization. For a pure collinear magnetic reflection in our case, we have $N = R_{ni} = J_{ni} = J_{ij} = 0$, reducing the matrix to the simplified form used in the main text.

### Least-Squares Fitting and $\chi^2(\beta, \theta)$

Given measured $p_{fi}^{(\mathrm{exp})}$ values for each reflection and theoretical $p_{fi}^{(\mathrm{model})}(\beta, \theta)$, we define a residual for each matrix element:

$$r_l(\beta, \theta) = \frac{p_{fi}^{(\mathrm{exp})} - p_{fi}^{(\mathrm{model})}(\beta, \theta)}{\sigma p_{fi}^{(\mathrm{exp})}}\,.$$

Here, $\sigma p_{fi}^{(\mathrm{exp})}$ denotes the experimental errors. Summing the squares,

$$\chi^2(\beta, \theta) = \sum_i \left[r_i(\beta, \theta)\right]^2,$$

gives the *weighted least-squares* cost function. Points with smaller $\sigma p_{ij}$ (higher precision) carry more weight in constraining the fit. We use MATLAB's `lsqcurvefit` routine, which iteratively tries different $(\beta, \theta)$ values, evaluates the residual vector $\mathbf{r}$, and adjusts the parameters to reduce $\sum_i r_i^2$.

After converging on the best fit $(\beta_{\mathrm{best}}, \theta_{\mathrm{best}})$ we obtain the Jacobian $J$. In a standard least-squares approximation (near the minimum), $\mathrm{Cov}(\beta, \theta) \approx (J^\top J)^{-1}$, and the standard error ($1\sigma$) on $\beta$ and $\theta$ is given by the square root of the diagonal elements of this covariance matrix: $\sigma(\beta) = \sqrt{\mathrm{Cov}_{\beta\beta}}$ and $\sigma(\theta) = \sqrt{\mathrm{Cov}_{\theta\theta}}$. An alternative scenario sets $\beta = 0$. We then minimize $\chi^2(0, \theta)$ over $\theta$ using the same routine but with $[\beta = 0, \theta]$.

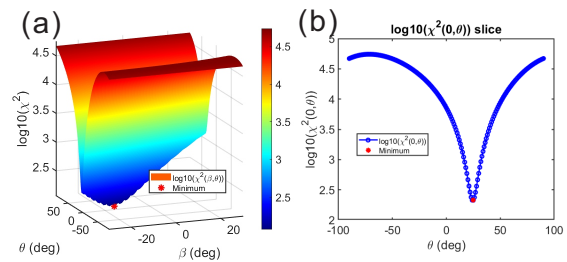

FIG. 2. (a) Three-dimensional surface plot of $\chi^2(\beta, \theta)$ with its global minimum marked in red. The best-fit parameters are $\beta = -13.8 \pm 1.5°$ and $\theta = 15.7 \pm 1.1°$. The resulting values are $\chi^2 = 116.6$, degrees of freedom (dof) = 34, and $\chi^2/\mathrm{dof} = 3.43$. (b) Calculated $\chi^2(0, \theta)$ as a function of $\theta$ assuming $\beta = 0$ with its minimum marked in red. The best-fit angle is $\theta = 24.5(3)°$, yielding a minimum $\chi^2$ of 209.7. With 35 degrees of freedom (dof = 35), the reduced chi-square is $\chi^2/\mathrm{dof} = 5.99$.

TABLE II. Comparison of refined $\chi^2(\beta, \theta)$ values for two other magnetic structure models with different $\beta$ and $\theta$. The negative sign of $\beta$ indicates an opposite rotation direction compared to the assumed convention.

| Models | This work | II[11] | III[5] |
|---|---|---|---|
| $\beta$ (°) | $-13.8(1.5)$ | 0 | 0 |
| $\theta$ (°) | 15.7(1.1) | 32 | 35 |
| $\chi^2(\beta, \theta)$ | 117 | 925 | 7441 |

### Analysis of the Constrained Model with $\beta = 0°$

To test if the magnetic structure is confined to the *ac*-plane, we fixed the in-plane twist to $\beta = 0°$ and allowed the out-of-plane tilt to freely optimize. This yields a model with $\theta \approx 24.5°$, but, the goodness-of-fit severely worsens to $\chi^2 = 210$, compared to $\chi^2 = 117$ for our freely refined model. We plot the calculated polarization matrix elements $P_{ij}$ of this constrained model against the experimental data in Fig. 3. In stark contrast to the excellent agreement achieved by our freely refined model (shown in Fig. 3(c) of the main text), this model fails to reproduce the experimental data, in particular, on the observed off-diagonal $P_{yz}$ and $P_{zy}$ components. This provides a clear visual confirmation of why the goodness-of-fit degrades so drastically without the in-plane twist.

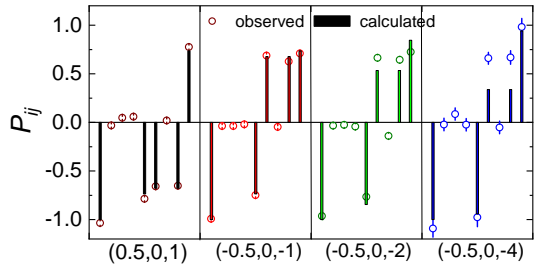

FIG. 3. Comparison of the observed (open circles) and calculated (solid bars) spherical neutron polarization matrix elements $P_{ij}$ for the artificially constrained magnetic structure model ($\beta = 0°$, $\theta \approx 24.5°$). For each reflection, the nine elements are plotted in the sequence $P_{xx}, P_{xy}, P_{xz}, P_{yx}, P_{yy}, P_{yz}, P_{zx}, P_{zy}, P_{zz}$. The clear discrepancy between the measured data and the calculations, particularly for the off-diagonal $P_{yz}$ and $P_{zy}$ components, visually demonstrates the inadequacy of an $ac$-plane confined moment mode to describe the magnetic structure.

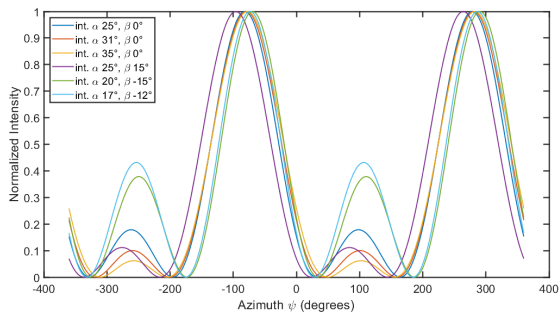

FIG. 4. Simulated azimuthal $\Psi$ dependence of the normalized REXS intensity for the magnetic reflection $(0, 1, 1.67)$ (indexed in the pseudo-orthorhombic notation that is used in the REXS experiments [11, 12]) corresponding to different magnetic structure models, parameterized by $(\beta, \theta) = (0°, 35°), (0°, 31°), (0°, 25°), (-12°, 17°)$ and others.

## REXS AZIMUTHAL SCAN: SIMULATION AND DISCUSSION

We simulated the REXS azimuthal scan for various magnetic structure models, including Model II obtained via the REXS experiments [11, 12] and the model obtained via our two polarized neutron diffraction experiments as well as some other models that are parameterized by $(\beta, \theta)$ (see Fig. 4). It clearly shows that introducing nonzero $\beta$ has actually a significant effect on the REXS azimuthal scan, which in principle should have been observable if it had been present in the samples measured.

Methodologically, REXS and unpolarized neutron diffraction critically rely on a very accurate and reliable measurement of the absolute intensity of magnetic reflections. Because these techniques measure the sum of the squared magnetic components and are inherently coupled to unknown domain volume fractions and scale factors, they are prone to discrete orientational, multi-domain and crystalline-mosaicity ambiguities. Furthermore, earlier analyses assumed a $C2/m$ symmetry, artificially constraining the moment to the $ac$ plane ($\beta = 0°$). In contrast, SNP directly measures the cross-correlations between the orthogonal magnetic components as domain-independent ratios. This off-diagonal sensitivity breaks the degeneracy, unambiguously locking in both the tilt and twist.

Having mentioned the unique strength of polarized neutron scattering, it is still difficult to explain why this 3D orientation of the ordered magnetic moment was not captured by the REXS measurements on single-domain samples [11, 12]. This could suggest that the magnetic structure model obtained in this study via polarized neutron scattering is not compatible with the REXS model. There are various possible explanations for this discrepancy. Sample dependence among different crystal growth sources may still be relevant. Furthermore, REXS is a near-surface probe ($\sim 1$–$2\,\mu$m depth) susceptible to cleavage-induced strain that can locally alter anisotropic exchange interactions, whereas our neutron experiments probe the pristine bulk of a stacking-fault-minimized crystal. Given that the nature of the strongly spin-orbit entangled electronic states in $\alpha$-RuCl$_3$ is sensitively dependent on, such as the local crystal-electric field of trigonal symmetry, and the presence of possible strong covalency due to Ru$^{3+}$-Cl$^-$ hybridization [13], a more interesting and plausible possibility, as also mentioned in the main text, is that the magnetic moment seen by REXS may differ from that measured by neutrons [14]. In this regard, a further understanding of the REXS cross section applied to this compound from the theory side would be necessary.


	
\end{document}